\documentclass[pdflatex,sn-mathphys]{sn-jnl}% Math and Physical Sciences Reference Style

\jyear{2022}%

\begin{document}

\title[Segmentation of the Great Arteries for CFD]{Automatic Segmentation of the Great Arteries for Computational Hemodynamic Assessment}


%%==================================%%
%% authors                          %%
%%==================================%%

\author[1]{\fnm{Javier} \sur{Montalt-Tordera}}\email{javier.montalt@ucl.ac.uk}
\equalcont{These authors contributed equally to this work.}

\author[1]{\fnm{Endrit} \sur{Pajaziti}}\email{endrit.pajaziti.13@ucl.ac.uk}
\equalcont{These authors contributed equally to this work.}

\author[2]{\fnm{Rod} \sur{Jones}}\email{rod.jones@gosh.nhs.uk}

\author[1]{\fnm{Emilie} \sur{Sauvage}}\email{e.sauvage@ucl.ac.uk}

\author[3,4]{\fnm{Rajesh} \sur{Puranik}}\email{rajesh.puranik@sydney.edu.au}

\author[3,4]{\fnm{Aakansha Ajay Vir} \sur{Singh}}\email{AakanshaAjayVir.Singh@health.nsw.gov.au}

\author[1]{\fnm{Claudio} \sur{Capelli}}\email{c.capelli@ucl.ac.uk}

\author[1]{\fnm{Jennifer} \sur{Steeden}}\email{jennifer.steeden@ucl.ac.uk}

\author[1]{\fnm{Silvia} \sur{Schievano}}\email{s.schievano@ucl.ac.uk}

\author*[1]{\fnm{Vivek} \sur{Muthurangu}}\email{v.muthurangu@ucl.ac.uk}

%%==================================%%
%% affiliations                     %%
%%==================================%%

\affil*[1]{\orgdiv{UCL Institute of Cardiovascular Science}, \orgname{University College London}, \orgaddress{\city{London}, \country{United Kingdom}}}

\affil[2]{\orgname{Great Ormond Street Hospital}, \orgaddress{\city{London}, \country{United Kingdom}}}

\affil[3]{\orgname{Children’s Hospital at Westmead}, \orgaddress{\city{Sydney}, \country{Australia}}}

\affil[4]{\orgdiv{Faculty of Medicine and Health}, \orgname{University of Sydney}, \orgaddress{\city{Sydney}, \country{Australia}}}

%%==================================%%
%% abstract                         %%
%%==================================%%

\abstract{\textbf{Background:} Computational fluid dynamics (CFD) is increasingly used for the assessment of blood flow conditions in patients with congenital heart disease (CHD). This requires patient-specific anatomy, typically obtained from segmented 3D cardiovascular magnetic resonance (CMR) images. However, segmentation is time-consuming and requires expert input. This study aims to develop and validate a machine learning (ML) method for segmentation of the aorta and pulmonary arteries for CFD studies.

\textbf{Methods:} 90 CHD patients were retrospectively selected for this study. 3D CMR images were manually segmented to obtain ground-truth (GT) background, aorta and pulmonary artery labels. These were used to train and optimize a U-Net model, using a 70-10-10 train-validation-test split. Segmentation performance was primarily evaluated using Dice score. CFD simulations were set up from GT and ML segmentations using a semi-automatic meshing and simulation pipeline. Mean pressure and velocity fields across 99 planes along the vessel centrelines were extracted, and a mean average percentage error (MAPE) was calculated for each vessel pair (ML vs GT). A secondary observer (SO) segmented the test dataset for assessment of inter-observer variability. Friedman tests were used to compare ML vs GT, SO vs GT and ML vs SO metrics, and pressure / velocity field errors.

\textbf{Results:} The network’s Dice score (ML vs GT) was 0.945 (interquartile range: 0.929–0.955) for the aorta and 0.885 (0.851–0.899) for the pulmonary arteries. Differences with the inter-observer Dice score (SO vs GT) and ML vs SO Dice scores were not statistically significant for either aorta or pulmonary arteries ($p = 0.741$, $p = 0.061$). The ML vs GT MAPEs for pressure and velocity in the aorta were 10.1\% (8.5–15.7\%) and 4.1\% (3.1–6.9\%) respectively, and for the pulmonary arteries 14.6\% (11.5–23.2\%) and 6.3\% (4.3–7.9\%), respectively. Inter-observer (SO vs GT) and ML vs SO pressure and velocity MAPEs were of a similar magnitude to ML vs GT ($p > 0.2$).

\textbf{Conclusions:} ML can successfully segment the great vessels for CFD, with errors similar to inter-observer variability. This fast, automatic method reduces the time and effort needed for CFD analysis, making it more attractive for routine clinical use.
}


\maketitle

\section{Background}\label{background}

The past two decades have seen increasing interest in the use of computational fluid dynamics (CFD) for the assessment of cardiovascular disease, including congenital heart disease (CHD) \cite{Biglino2017ComputationalTranslation}. Computational fluid dynamics models enable realistic calculation of patient-specific blood flow conditions and provide valuable insights into pathological hemodynamics. These models can also be used to predict hemodynamic response to interventions, thereby aiding therapeutic planning.

The patient specific anatomies needed for CFD are often derived from three dimensional (3D) cardiovascular magnetic resonance (CMR), particularly cardiac and respiratory gated whole-heart sequences. This is because whole-heart images have the sharp borders and high contrast necessary for semi-automated segmentation of cardiovascular structures. 

Automatic or semi-automatic segmentation methods based on shape models or level-set algorithms have existed for years \cite{Wang2014FastPropagation,Weese2001ShapeSegmentation,Peters2007AutomaticVolumes}. However, they often require some type of user input or manual post-correction, rely on priors which are not readily available, or may struggle to adapt to abnormal anatomies (e.g., CHD). Thus, segmentation remains one of the most user-intensive and time-consuming parts of the CFD workflow, and one of the barriers to greater clinical use.

Recently, it has been shown that machine learning (ML) can accurately segment ventricles and great vessels from CMR images \cite{Chen2020DeepReview,Berhane2020FullyLearning,Zhuang2019EvaluationChallenge}. Quantitative metrics derived from ML segmentations (e.g., ventricular volumes) compare well with manual segmentations \cite{Bai2018AutomatedNetworks,Bottcher2020FullyAlgorithm} and these techniques are now entering clinical practice. However, the effectiveness of ML segmentation for CFD has not previously been investigated. 

The aims of this study were to: (i) develop a ML method for simultaneous segmentation of the aorta and pulmonary arteries from whole heart CMR images in patients with paediatric or adult CHD, (ii) compare conventional and ML segmentations using traditional image-based scores, (iii) compare CFD metrics derived from both conventional and ML segmentations, and (iv) investigate the association between image-based scores and CFD errors.

\section{Methods}\label{methods}

\subsection{Subjects}\label{subjects}

90 cardiac triggered, respiratory navigated, 3D whole heart, balanced, steady state, free precession (WH-bSSFP) data were collected from previously scanned children and adults with paediatric or congenital heart disease (excluding patients with single ventricles). All patients were scanned on a 1.5T scanner (Avanto, Siemens Healthineers AG, Erlangen, Germany) using a standard WH-bSSFP sequence \cite{Steeden2020RapidSuper-resolution}. The imaging protocol was as follows: orientation: sagittal, matrix size: 256 x 144 x 96 (head-foot, anterior-posterior, left-right), acquired voxel size: 1.6 mm (isotropic), flip angle: 90°. Image acquisition was accelerated using GRAPPA (factor of 2 along phase encoding dimension) and partial Fourier (factor of 6/8 along both phase and slice encoding dimensions). The use of retrospectively collected training and test data was approved by the local research ethics committee, and written consent was obtained from all subjects/guardians (Ref: 06/Q0508/124).

Additionally, 10 external examples were retrospectively collected from a different centre. These were scanned on a 1.5 T scanner (Ingenia, Philips Healthcare, Amsterdam, Netherlands) with the following imaging protocol: orientation: axial, matrix size: 240 x 240 x 110 (left-right, anterior-posterior, head-foot), acquired voxel size: 1.44 mm (isotropic), flip angle: 90°. Image acquisition was accelerated using SENSE (reduction factor of 2) and partial Fourier (6/8). Collection of this data and sharing with our site was approved by the local research ethics committee, and written consent was obtained from all subjects/guardians (Protocol No. X20-0237 and 2020/ETH01333).

\subsection{Ground Truth Segmentation}\label{gtseg}

Reference standard conventional segmentation of the aorta and pulmonary arteries was performed using a semi-automatic technique with manual correction (Plug-ins built in Horos v4.0, Horosproject.org sponsored by Nimble Co LLC d/b/a Purview, Maryland, USA). Initial segmentation was done using the fast level-set method \cite{Wang2014FastPropagation}. This requires the user to: (i) set a threshold, (ii) place seeds in the vessel of interest and (iii) add blocking regions to prevent segmentation of unwanted structures. The quality of this initial segmentation is dependent on both the underlying anatomy and the image quality, but manual correction is always required to remove unwanted structures and clip vessels. The proximal limit of both the aortic and pulmonary artery segmentations was the semi-lunar valve. The distal limit of the segmentations were the diaphragmatic level of the aorta, and hilar branches of the pulmonary arteries. Head and neck arteries were manually removed at their origin.

All 90 datasets were segmented by a primary observer (RJ — 10 years’ experience in CMR post-processing). We refer to the primary observer’s segmentations as the ground truth (GT). In addition, a secondary observer (VM — 19 years’ experience in CMR post-processing) segmented 10 of these images (test set, see below) to investigate inter-observer variability. We refer to these as the secondary observer (SO) data. The 10 external examples were segmented by the secondary observer only following the same procedure.

\subsection{Data Preparation}\label{dataprep}

Prior to ML training, the pixel intensities of the WH-bSSFP data were normalized (range [0, 1]). The aortic and pulmonary binary segmentation masks were concatenated in the channel dimension and combined with a third channel containing a binary “background” mask (one-hot encoding). The images and 3-channel segmentation masks were either centrally cropped or symmetrically zero-padded to a fixed matrix size of 160×96×64 (superior-inferior, anterior-posterior, left-right). Finally, the image-label pairs were randomly split into a training set (70 examples, 78\%), a validation set (10 examples, 11\%) and a test set (10 examples, 11\%). This split was used to maximize the size of the training set, while providing sufficient data for validation and statistical analysis.

The examples from the external test set were reoriented, interpolated and cropped to match the orientation, matrix size and voxel size of the in-house data.

\subsection{Network Architecture}\label{netarch}

\begin{figure}[htbp]
\centering
\includegraphics[width=0.9\textwidth]{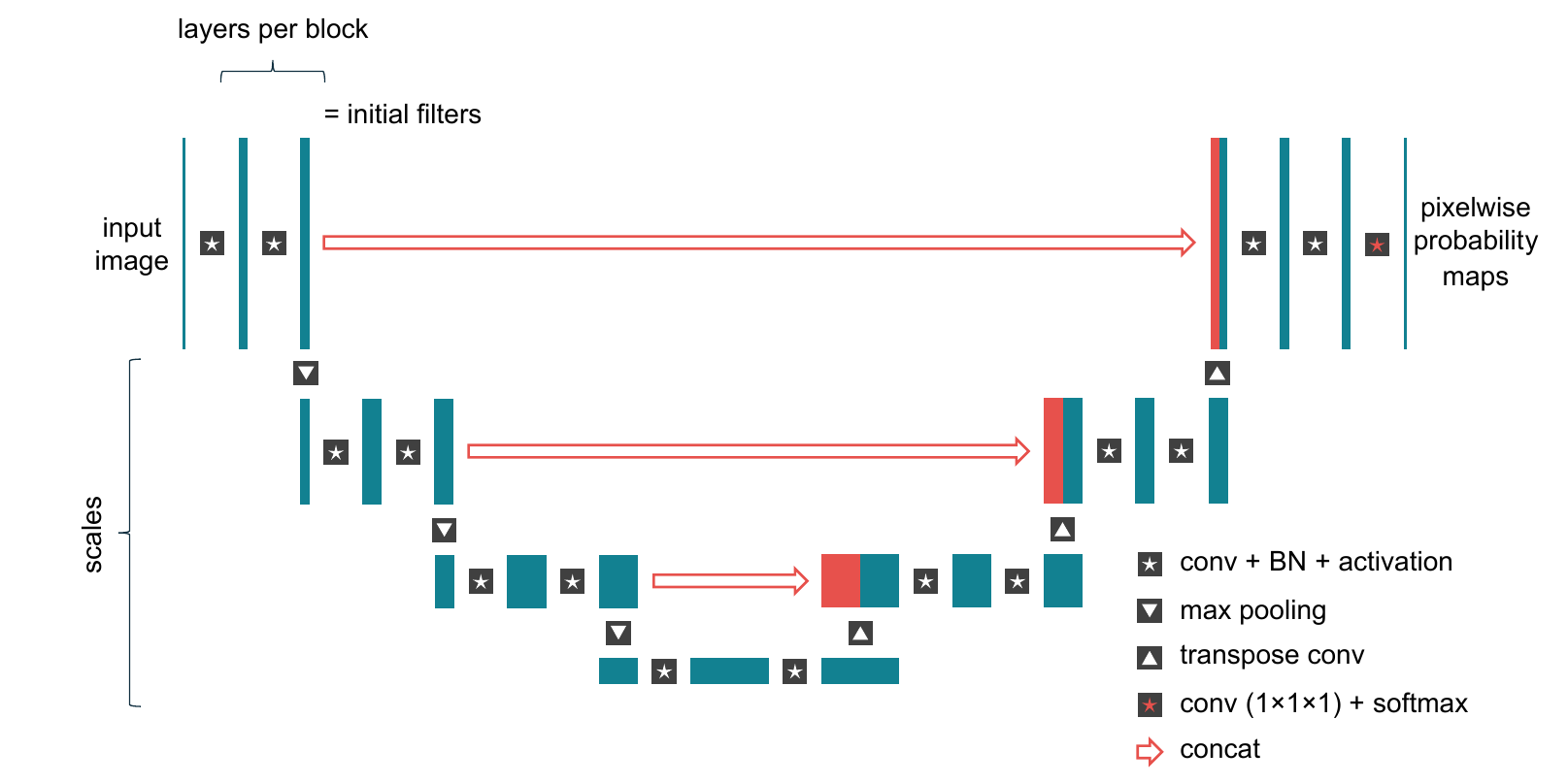}
\caption{Parameterized network architecture. The height of the blocks represents changes in spatial resolution while the width represents the number of filters or channels. The figure shows the three network parameters whose value was optimized: initial filters, layers per block and scales. This example is shown with 2 layers per block and 3 scales. BN: batch normalization.}\label{fig1}
\end{figure}

A U-Net \cite{Ronneberger2015U-Net:Segmentation} convolutional neural network was used to simultaneously segment the aorta and pulmonary arteries from WH-bSSFP data. The network architecture is shown in Fig. \ref{fig1}. Each convolutional layer was followed by a batch normalization layer and a rectified linear unit (ReLU) activation. Downscaling was performed using max-pooling layers and upscaling was performed using transpose convolution layers. The number of convolutional filters after the first layer was set to double after each downscaling layer and halve after each upscaling step. The final convolutional layer has three filters (equalling the number of possible classes – aorta, pulmonary artery and background), followed by a softmax activation. Final predicted labels were obtained by assigning each pixel to the class with the highest probability.

\subsection{Training, Hyperparameter Optimization and Evaluation}\label{traineval}

The network implementation and training scheme were parametrized to allow investigation of multiple hyperparameter values, with the full search space shown in Table \ref{hyperparameter-search-space}. We used the Hyperband algorithm \cite{Li2018Hyperband:Optimization} to perform efficient hyperparameter optimization. This method samples the search space randomly and adaptively allocates more computational resources to the most promising hyperparameters combinations. Dice score was used by the Hyperband to assess performance and choose the final model.

\begin{table}[htbp]
\centering
\caption{Hyperparameter search space.}\label{hyperparameter-search-space}
\begin{tabular}{@{}lll@{}}
\toprule
\textbf{Parameter} & \textbf{Type} & \textbf{Domain} \\
\midrule
Scales           & Architecture & \{2, 3, 4\}                                \\
Layers per block & Architecture & \{2, 3, 4\}                                \\
Initial filters  & Architecture & \{32, 64\}                                 \\
Learning rate    & Training     & {[}0.0001, 0.01{]}\footnotemark[1]                         \\
Batch size       & Training     & \{2, 4\}                                   \\
Loss function    & Training     & \{CCE\footnotemark[2], Dice, IoU, Tversky, focal Tversky\} \\
\botrule
\end{tabular}
\footnotetext{$\{ \cdot \}$  represents a discrete set of values; $ [ \cdot ] $ represents a continuous interval of real values.}
\footnotetext[1]{The learning rate was sampled using a log-uniform distribution.}
\footnotetext[2]{CCE: categorical cross-entropy.}
\end{table}

The neural network and related functionality were implemented and trained using TensorFlow \cite{Abadi2016TensorFlow}. In particular, the model implementation, losses and metrics are available in TensorFlow MRI \cite{TensorFlowMRI}, an open-source framework developed in-house. Weights were initialized using He’s method \cite{He2015DelvingClassification} and optimized using the Adam algorithm \cite{Kingma2015Adam:Optimization}. Training, including hyperparameter optimization, took ~24 hr on an Nvidia Titan RTX GPU with 24 GB of onboard RAM (Nvidia Corporation, Santa Clara, CA, USA).

The optimized ML model was evaluated on the test dataset against the ground truth segmentations (ML vs GT). The accuracy of segmentation was quantified using several image-based segmentation metrics: Dice score, Intersection over Union (IoU), Hausdorff distance (HD) and average surface distance (ASD). Each metric was computed independently for each vessel. Additionally, the same metrics were calculated for the secondary observer’s segmentation against the GT (SO vs GT), and between the ML model and the secondary observer (ML vs SO).

To assess generalization ability to data from other sources, the ML model was also evaluated on the external test set. We computed the same set of metrics (Dice, IoU, HD and ASD) between the ML predictions (Ext-ML) and the ground truth segmentations (we refer to these as Ext-SO, because the manual segmentation was performed by the same person as the SO data).

Prior to evaluation, ML masks were filtered to remove all but the largest connected component, as identified using 3D connected component labelling with 26-connectivity \cite{Silversmith2021Cc3d:Images}. This postprocessing step was used to eliminate small background regions which had been misclassified as vessels.

\subsection{Surface and Volume Meshing Pipeline}\label{mesh}

\begin{figure}[htbp]
\centering
\includegraphics[width=0.9\textwidth]{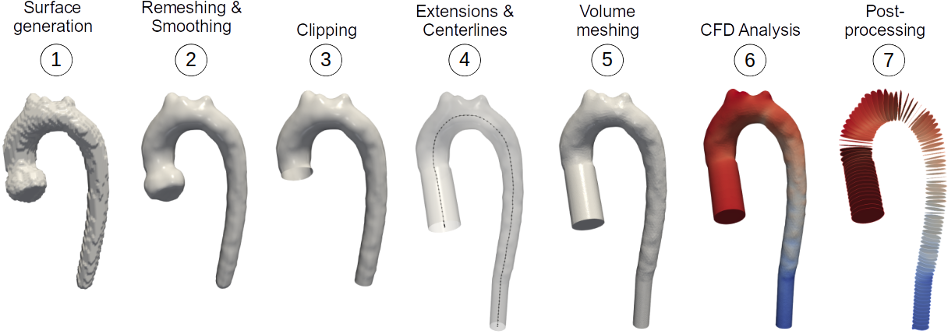}
\caption{Automatic mesh processing pipeline from segmentation to CFD analysis, followed by post-processing to reshape the data in a consistent format between subjects (99 planes from inlet to outlet containing average pressure and velocity).}\label{fig2}
\end{figure}

The resultant segmentation masks were converted into finite element volume meshes, using the processes shown in Fig. \ref{fig2}. The masks (GT, ML and SO) were first transformed into surface meshes, by applying the marching cubes algorithm (implemented in VMTK), and then re-meshed and smoothed with consistent parameters. The surface meshes were clipped manually at inlets and outlets to create planar surfaces; 30 mm extensions were added to ensure the fluid flow in the region of interest was fully developed and capped at the ends to generate close surfaces. These were meshed with tetrahedral elements to build the final unstructured grid for CFD analysis. The grid resolution was determined through a sensitivity analysis (see Appendix \ref{appA}).

To assess the effect of the manual clipping of the anatomies on the CFD, the simulations from the ML segmentations were also re-run using the same inlet and outlet locations as the GT data, defined by overlaying the GT clipped geometry to the corresponding ML geometry.

\subsection{Computational Fluid Dynamics and Boundary Conditions}\label{cfd}

CFD simulations were carried out using the solver Fluent 19.0 (Ansys, Canonsburg, PA, USA). Blood was modelled as an incompressible Newtonian fluid with density 1060 kg/m3 and 0.004 Pa$\cdot$s dynamic viscosity \cite{Gijsen1999TheModel}. Vessel walls were considered rigid, and no-slip conditions were imposed. A laminar, steady-state model was selected to simulate blood flow at peak systole \cite{Caballero2013AAorta,Bonfanti2017ComputationalData}. A generalisable inlet condition for the aorta and pulmonary artery was applied to all subjects with a uniform (plug) inlet velocity profile of 0.66 m/s for the aorta and 0.57 m/s for the pulmonary artery \cite{Gabe1969MeasurementProbe}. The outlets for all cases were assumed to be at zero pressure and the convergence criteria was set at $10^{-4}$ for the residual errors. Simulations were run on a Dell workstation, with a Xeon CPU E5-2630 (24 processors at 2.3 GHz), 32 GB RAM and an Nvidia GeForce GTX 1080 Ti.

\subsection{CFD Post-Processing and Analysis}\label{cfdpost}

To compare the flow field between the pairs of different unstructured meshes (ML vs GT, SO vs GT and ML vs SO), correspondence was created by subdividing each vessel with 99 planes orthogonal to the centrelines, calculated in VMTK, and equally distanced. Static pressure and velocity magnitude were averaged in each plane (see Fig. \ref{fig2}) and a percentage error was calculated for each plane pair, using ML as reference (GT in the comparison between SO and GT). The mean absolute percentage errors (MAPE) for pressure and velocity were computed for each vessel pair.

\subsection{Statistical Analysis}\label{stats}

Shapiro-Wilk tests were used to test the normality of the different segmentation metrics and CFD errors, grouped by vessel (aorta and pulmonary artery), and segmentation pair (ML vs GT, ML vs SO and SO vs GT). Wilcoxon signed rank tests were used to compare the pressure and velocity errors for the ML vs GT group. Mann-Whitney U-tests were used to compare segmentation metrics and flow field errors between the aorta and the pulmonary artery, for the ML vs GT group. Friedman tests for repeated measurements were performed to compare segmentation metrics and flow field errors between the ML vs GT, ML vs SO and SO vs GT groups, for both aorta and pulmonary artery segmentations. Significant Friedman test results were followed up by pairwise Wilcoxon post-hoc tests. Additionally, Wilcoxon signed rank tests were used to compare ML vs GT and SO vs GT metrics for both aorta and pulmonary artery segmentations. Mann-Whitney U-tests were used to compare Ext-ML vs Ext-SO segmentation metrics against ML vs SO metrics. Wilcoxon signed rank tests were used to compare the pressure and velocity errors for the manually clipped ML vs GT data and the equally clipped ML vs GT data. Pearson’s correlation coefficient was used to measure the linear relationship between each pair of a segmentation metric (i.e., Dice, IoU, HD or ASD) and a flow field error (pressure or velocity MAPEs), for both aorta and pulmonary artery segmentations. The \emph{p}-value was calculated for each comparison to test non-correlation. Throughout this work, a \emph{p}-value $<$ 0.05 was considered statistically significant.

\section{Results}\label{results}

\subsection{Hyperparameter Optimization}\label{results-hparams}

A total of 124 hyperparameter configurations were sampled during the neural network optimization procedure (see Appendix \ref{appB}). The best performing configuration was as follows: scales = 3, layers per block = 2, initial filters = 64, learning rate = 3.46x10-4, batch size = 2, and loss function = focal Tversky. This model was selected and used in all further experiments.

\subsection{ML Segmentation}\label{results-segmentation}

\begin{figure}[htbp]
\centering
\includegraphics[width=0.9\textwidth]{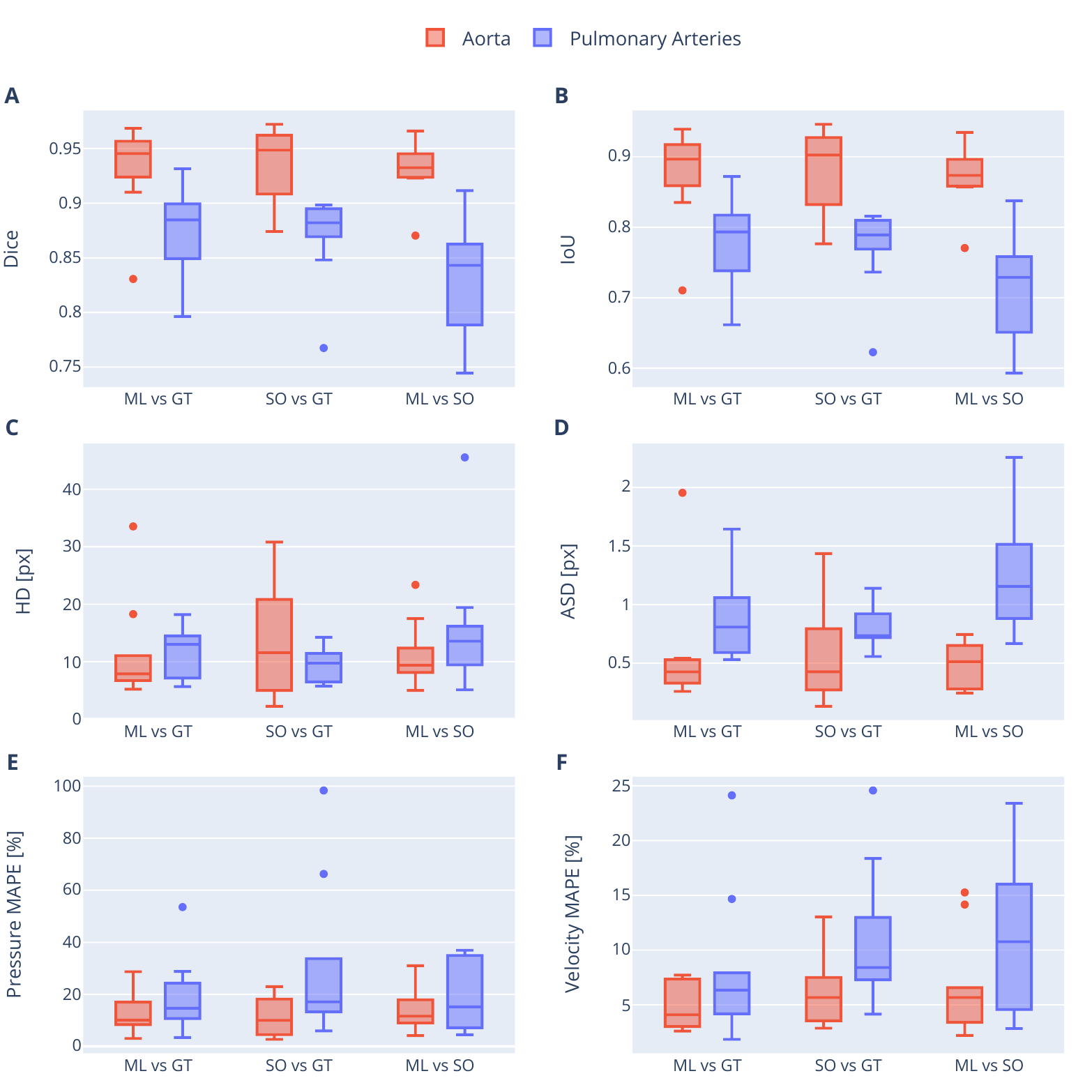}
\caption{Segmentation metrics and flow field errors. Three segmentations are compared in a pairwise fashion: machine learning (ML), ground truth (GT) and secondary observer (SO). (A–B): Confusion-based similarity metrics: Dice score and IoU. (C–D) Distance-based similarity metrics: Hausdorff distance (HD) and average surface distance (ASD), measured in pixels. (E–F) CFD-derived pressure and velocity mean average percentage errors (MAPE).}\label{fig3}
\end{figure}

The ML segmentation was successful in all 10 test datasets. The specific diagnoses for these patients were: repaired tetralogy of Fallot ($n = 1$), repaired Tetralogy of Fallot with mild right pulmonary artery stenosis ($n = 1$), Marfan syndrome with dilated aorta ($n = 1$), Marfan syndrome with pectus excavatum ($n = 1$), dilated pulmonary artery ($n = 1$), bicuspid aortic valve with dilated aorta and unrepaired VSD ($n = 1$), repaired double outlet right ventricle with right sided arch ($n = 1$), unrepaired atrial septal defect ($n = 1$), aortic regurgitation with dilated aorta ($n = 1$), post Ross procedure with mechanical aortic valve ($n = 1$). Inference time for the ML model was approximately 160 ms for simultaneous segmentation of aorta and pulmonary arteries (compared to approximately 30 minutes for manual segmentation of aorta and pulmonary arteries). There was good agreement between the ML and GT segmentation with a median Dice score of 0.945 (interquartile range: 0.929–0.955) for the aorta and 0.885 (0.851–0.899) for the pulmonary arteries. The Dice score was significantly higher for the aorta than the pulmonary arteries ($p = 0.002$) with similar findings observed for IoU, HD and ASD (Fig. \ref{fig3}A–D).

\begin{figure}[htbp]
\centering
\includegraphics[width=0.9\textwidth]{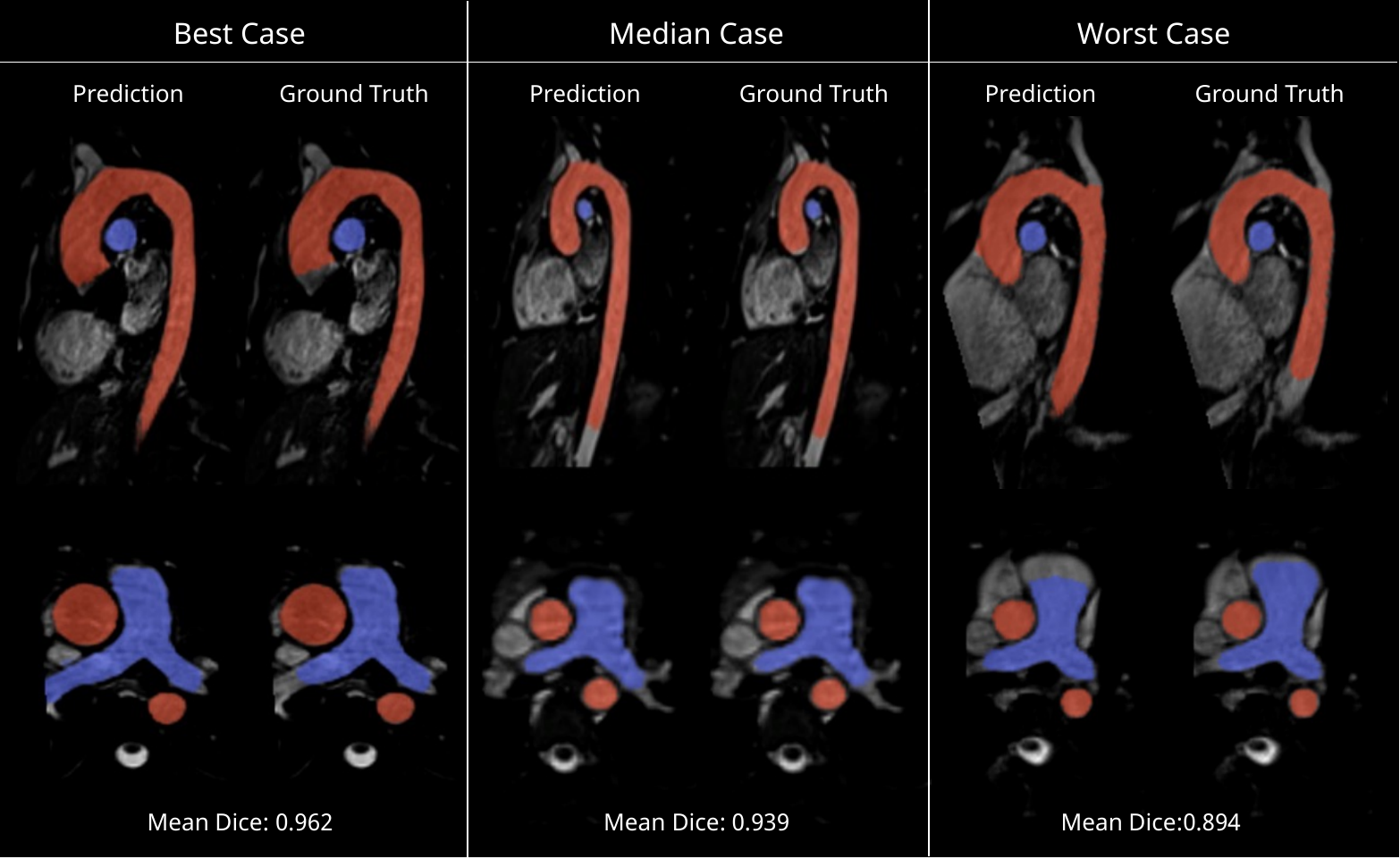}
\caption{Test set segmentation overlays. Predicted and ground truth masks are overlayed over the original images for the best, median and worst test cases. Aorta and pulmonary artery masks are shown in red and blue, respectively. Multiplanar reformats of the original 3D volume were manually selected on a case-by-case basis to be most informative. Best case had Ross procedure and mechanical aortic valve, the median case had an atrial septal defect and the worst case had a dilated pulmonary artery.}\label{fig4}
\end{figure}

The best, median and worst segmented images in terms of Dice score are shown in Fig. \ref{fig4}. The three main differences were: (i) the length of the vessel segmented, (ii) differences in pixel labelling that resulted in small deviations of the vessel border, and (iii) small protrusions at origin of the carotid and subclavian arteries in the ML segmentations of the aorta.

The aortic inter-observer Dice score (SO vs GT) was 0.949 (0.916–0.960) and was not significantly different from ML vs GT ($p = 0.575$). The pulmonary Dice score for the SO vs GT was 0.882 (0.870–0.894) and was also not significantly different from ML vs GT ($p = 0.721$). The ML vs SO Dice score was 0.933 (0.924–0.944) for the aorta, which was not significantly different from ML vs GT and SO vs GT ($p = 0.741$), and 0.843 (0.791–0.860) for the pulmonary arteries, which trended towards being lower than ML vs GT and SO vs GT ($p = 0.061$).

\begin{figure}[htbp]
\centering
\includegraphics[width=0.9\textwidth]{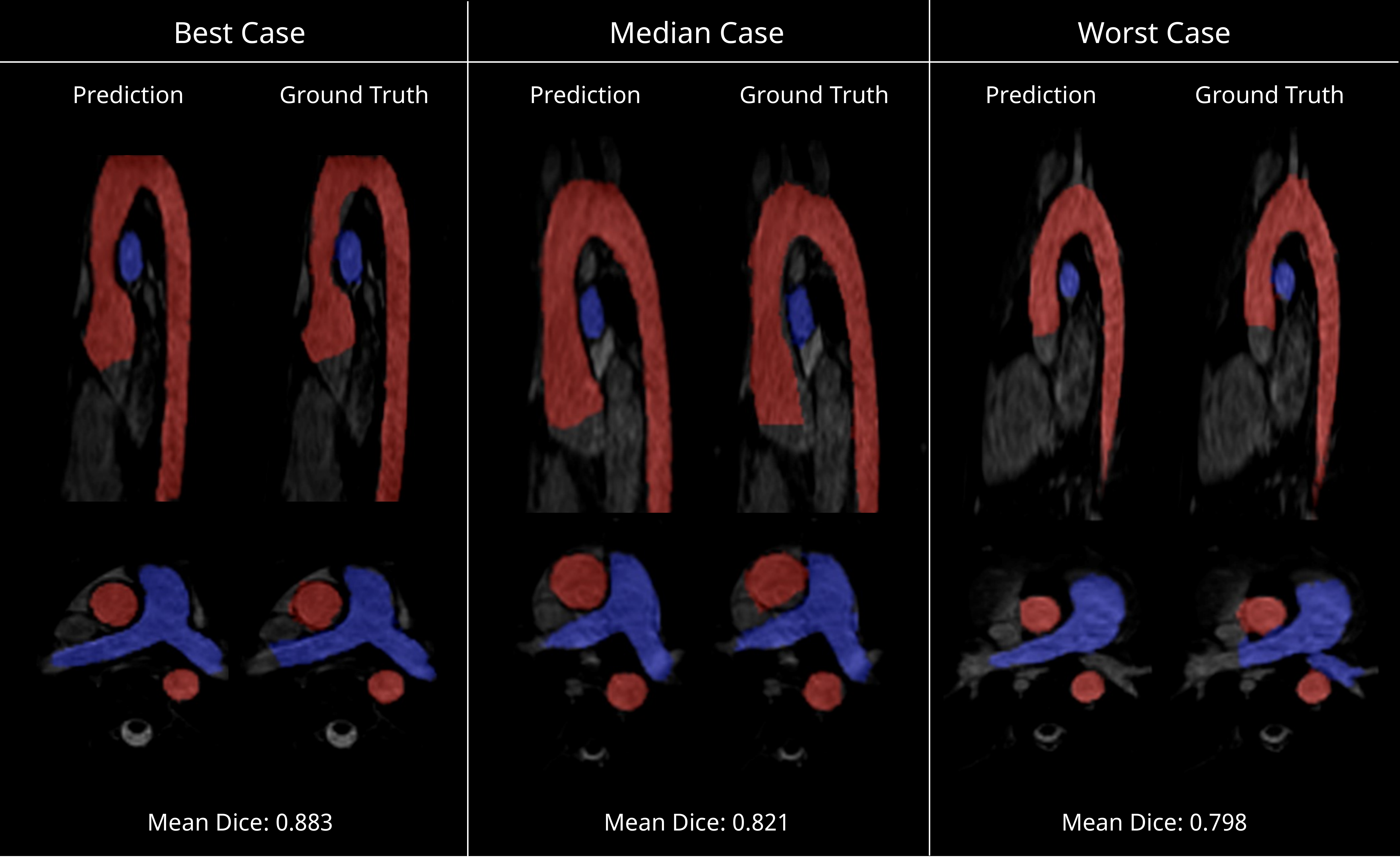}
\caption{External test set segmentation overlays. Predicted and ground truth masks are overlayed over the original images for the best, median and worst test cases. Aorta and pulmonary artery masks are shown in red and blue, respectively. Multiplanar reformats of the original 3D volume were manually selected on a case-by-case basis to be most informative. Best case had normal anatomy with gothic arch, the median case had bicuspid aortic valve and the worst case had mild left pulmonary artery stenosis.}\label{fig5}
\end{figure}

\begin{figure}[htbp]
\centering
\includegraphics[width=0.9\textwidth]{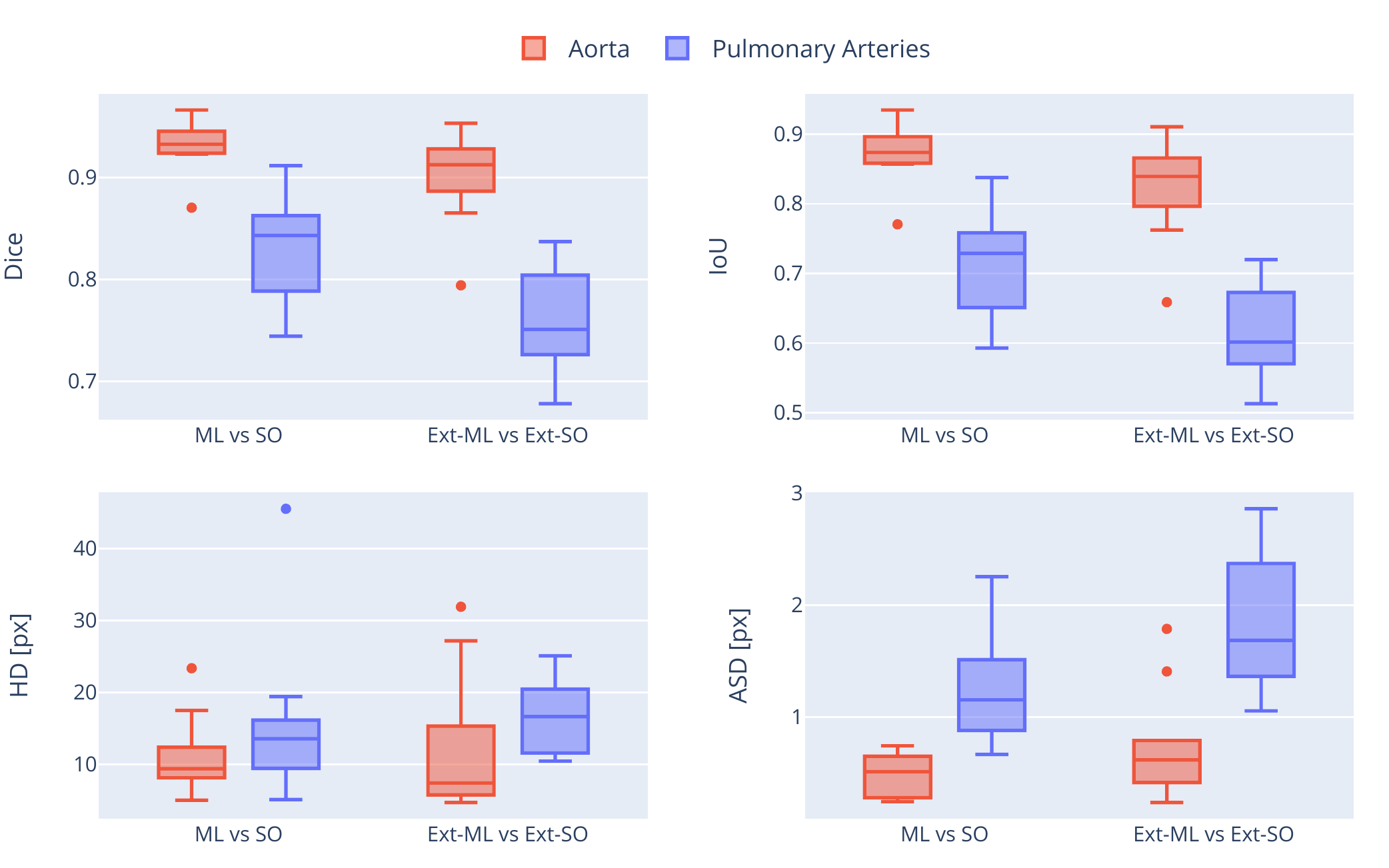}
\caption{Segmentation metrics. The model’s segmentation (ML) is compared with the an observer’s segmentation (SO) for two different datasets: our original data (ML vs SO) and an external test set from a different site and vendor (Ext-ML vs Ext-SO). (A–B): Confusion-based similarity metrics: Dice score and intersection-over-union (IoU). (C–D) Distance-based similarity metrics: Hausdorff distance (HD) and average surface distance (ASD), measured in pixels.}\label{fig6}
\end{figure}

The ML segmentation was also successful in the external dataset. The specific diagnoses for these patients were: cardiomyopathy ($n = 4$), normal anatomy ($n = 1$), repaired tetralogy of Fallot ($n = 1$), left pulmonary artery stenosis ($n = 1$), anomalous pulmonary venous drainage ($n = 1$), repaired coarctation of the aorta with hypoplastic arch ($n = 1$), bicuspid aortic valve with severe AR and dilated aortic root ($n = 1$). The best, median and worst examples from the external test set are shown in Fig. \ref{fig5}. There was reasonable agreement between the Ext-ML and Ext-SO segmentations, with a median Dice score of 0.913 (0.889–0.927) for the aorta and 0.751 (0.728–0.797) for the pulmonary arteries. Agreement was significantly lower than ML vs SO for the pulmonary arteries ($p = 0.011$), but not for the aorta ($p = 0.089$). Similar findings were observed for IoU, HD and ASD (Fig. \ref{fig6}).

\subsection{CFD Metrics}\label{results-cfd}

There was overall good agreement in CFD metrics calculated using ML and GT segmentations (Fig. \ref{fig3}E–F). The median MAPE for pressure and velocity in the aorta were 10.1\% (interquartile range: 8.5–15.7\%) and 4.1\% (3.1–6.9\%) respectively, and for the pulmonary arteries 14.6\% (11.5–23.2\%) and 6.3\% (4.3–7.9\%). Pulmonary artery MAPEs trended towards higher values compared to aortic MAPEs, but this did not reach statistical significance ($p = 0.081$ for pressure and $p = 0.093$ for velocity). However, pressure was more sensitive than velocity to different segmentations, with pressure MAPE being ~2.5x greater than velocity MAPE ($p < 0.001$).

\begin{figure}[htbp]
\centering
\includegraphics[width=0.9\textwidth]{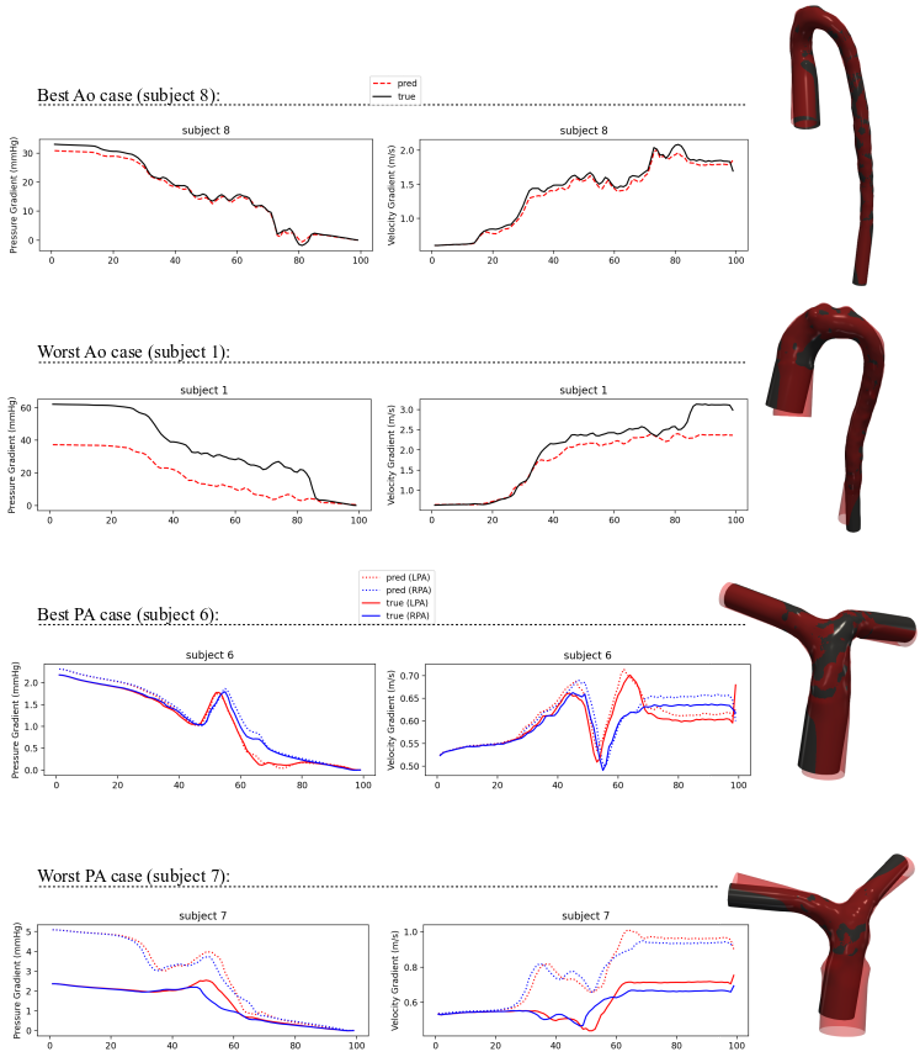}
\caption{Best and worst cases for aortas and PAs. Graphs show planar-averaged metrics along the length of the vessels, starting from the inlet. Black geometries correspond to ground truth, whereas red correspond to predictions. Best aorta case had atrial septal defect and the worst had a repaired double outlet right ventricle with right arch. The best PA has a bicuspid aortic valve with dilated aorta and unrepaired VSD and the worst is a Marfan syndrome with dilated aorta.}\label{fig7}
\end{figure}

\begin{figure}[htbp]
\centering
\includegraphics[width=0.9\textwidth]{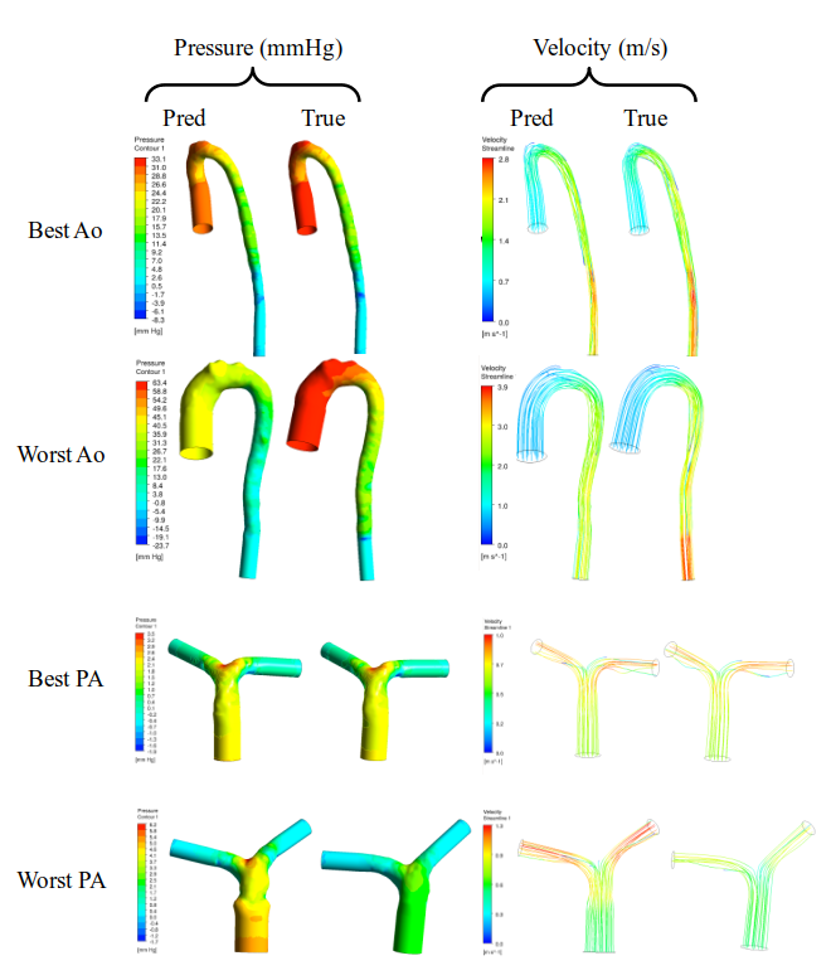}
\caption{Best and worst aorta and pulmonary artery predictions. Flow fields of pressure and velocity displayed.}\label{fig8}
\end{figure}

Fig. \ref{fig7} shows the surface meshes of test cases with the highest and lowest CFD MAPE, as well as pressure and velocities along the length of each vessel. Fig. \ref{fig8} shows pressure and velocity fields calculated using both ML and GT manual segmentations. The main difference in the surface meshes (particularly for the worst cases) were associated with the inlets and outlets (angle and size) and these differences propagated into pressure and velocity fields.

SO vs GT (inter-observer) and ML vs SO pressure and velocity MAPEs were of a similar magnitude to the errors from the ML segmentations (Fig. \ref{fig3}E–F, $p > 0.2$).  When the clipping planes of the GT segmentations were used on the ML geometries, the median pressure and velocity MAPEs were reduced to 8.0/3.1\% ($p < 0.01$) for the aorta, and to 10.4/3.7\% for the pulmonary artery ($p < 0.01$) (see Appendix \ref{appC}).

Fig. \ref{fig9} illustrates the relationship between the segmentation metrics and the CFD errors on the ML vs GT comparison. No statistically significant correlations were found between any of the metrics, either for the aorta or the pulmonary arteries, for either manual or equally clipped data.

\begin{figure}[htbp]
\centering
\includegraphics[width=0.9\textwidth]{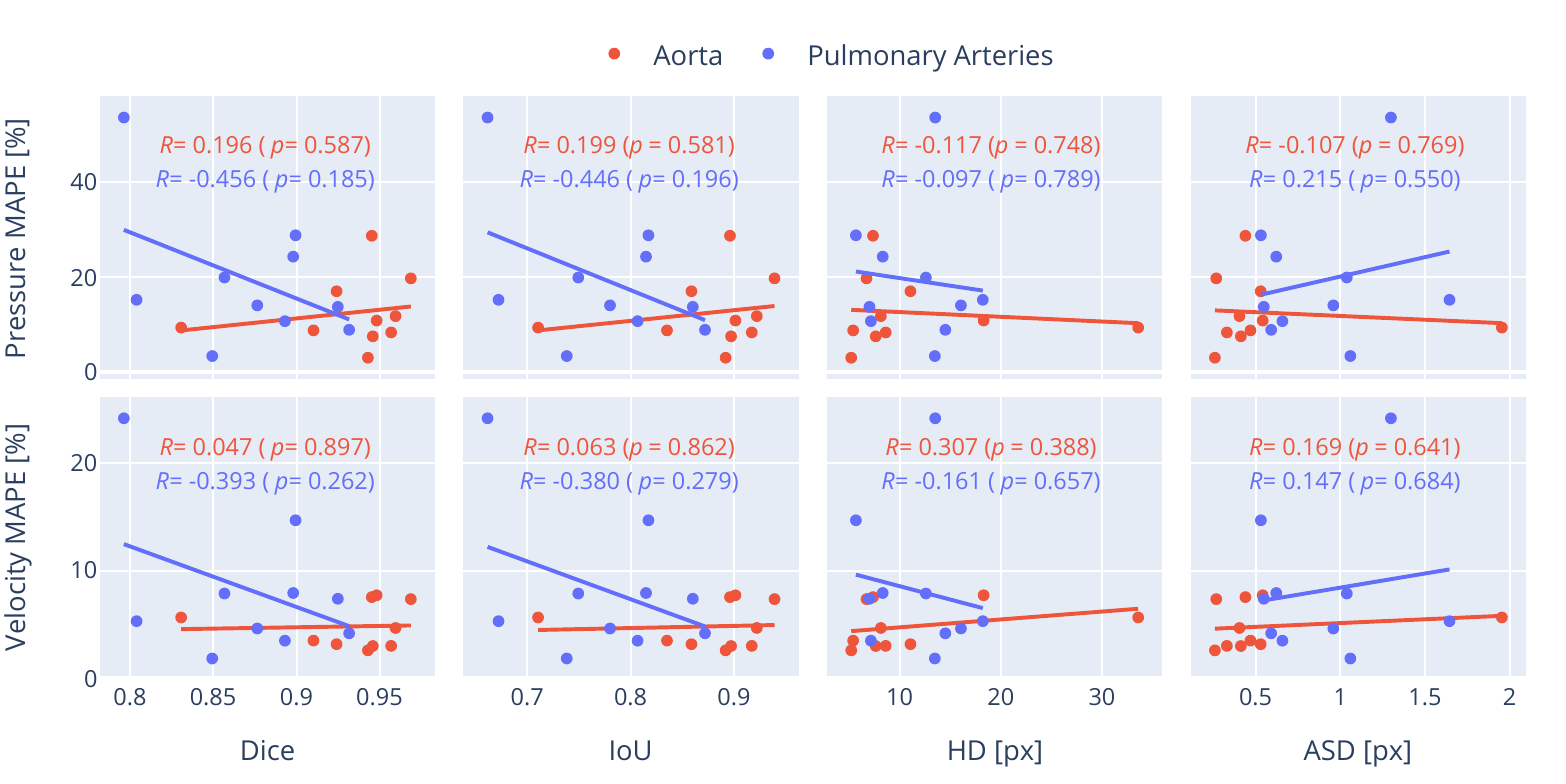}
\caption{Flow errors against similarity metrics. The figure shows a scatter-plot matrix where each point corresponds to a subject. In the abscissas, two confusion-based metrics, Dice and IoU, and two distance-based metrics, the Hausdorff distance and the average surface distance, measured in pixels. In the ordinates, the pressure and velocity mean average percentage errors (MAPE). All values are for the ML vs GT comparison. Red and blue colours identify aorta and pulmonary artery data, respectively. Trend lines are least-squares polynomial fits of degree 1. For Dice and IoU, higher is better (more similar). For Hausdorff distance, average surface distance and pressure and velocity MAPEs, lower is better.}\label{fig9}
\end{figure}
 
\section{Discussion}\label{discussion}

In this study, a deep neural network was trained to simultaneously segment the aorta and pulmonary arteries from 3D MRI data. As its primary purpose was to provide patient specific anatomies for CFD models, we evaluated accuracy using conventional image-based segmentation metrics and resulting errors in CFD measures. The main findings were: (i) the proposed network achieved high performance in terms of image-based segmentation metrics, (ii) there was reasonable agreement between CFD models derived from the ML and ground truth manual segmentation, (iii) these errors were similar in magnitude to those observed between two different manual segmentations, and (iv) there was no relationship between the segmentation metrics and the resulting CFD errors.

\subsection{ML Segmentation}\label{discussion-segmentation}

In data from the same distribution as the training data, the segmentation model achieved comparable or better performance than previously reported 3D ML segmentation of the great vessels, including in patients with congenital heart disease \cite{Berhane2020FullyLearning,Payer2017Multi-labelConfigurations}. This suggests that the chosen network architecture and subsequent hyperparameter optimization were sufficient for accurate segmentation. Nevertheless, there were some differences between the GT and ML segmentations and visual inspection reveals three main types of error. The first error was a tendency for ML to start and stop segmenting at slightly different points in the vessel compared to the ground truth. The second type of error was the presence of “bumps”, due to the segmentation masks bleeding out at the locations of arterial branches, particularly in the aorta. Both these errors can be considered failures to properly demarcate vessel limits, rather than failures to correctly label blood pool pixels. The third type of error was inaccurate labelling of blood pixels at the vessel border, resulting in subtle differences in surface geometry. It should be noted that none of these patients had very abnormal pulmonary vascular or aortic anatomy, which was necessary to ensure that CFD models could be created from the segmentations. However, further testing on complex congenital heart disease is necessary if segmentation models are to be used more widely. Extension to complex disease may require further enhancements, and several strategies could potentially help improve the ML segmentation accuracy and generalizability. These include increasing the amount and heterogeneity of training data, or performing data augmentation, both of which improve generalizability and performance of ML models \cite{Sun2017RevisitingEra,Krizhevsky2012ImageNetNetworks}. Another interesting option might be the inclusion of statistical shape models \cite{Bhalodia2018DeepSSM:Images,Raju2021DeepDelineation}, which could help ensure that the segmented shapes conform to common patterns.

The model was also tested on 3D data acquired on a different vendor scanner. Although the type of sequence (3D whole-heart bSSFP) and imaging protocol were similar to the original data, there were visually apparent differences in image quality and characteristics. Nevertheless, we observed reasonable segmentation quality. For the aorta, agreement with a human observer was only slightly lower than agreement with the same observer in our original data. However, there was a larger reduction in agreement for the pulmonary arteries. This suggests there is scope for improving the generalizability of the model. One of the best solutions for this is to include multi-site, multi-vendor data in the training set, but this would incur obvious labelling costs and potential data sharing difficulties. Other approaches to improve robustness to out-of-distribution data might be the use of data augmentation (e.g., domain translation methods to generate multi-vendor datasets \cite{Zhu2017UnpairedNetworks,Yan2019TheUnet-GAN}) and the use of strategies that incorporate additional domain knowledge \cite{Liu2021Anatomy-aidedReview}.

Of course, segmentation is a challenging task and we demonstrated that the agreement between two humans was similar to the agreement between ML and the GT human segmentation. This suggests that ML “errors” are approximately at the level of the inter-observer variability and similar observations have previously been made for aortic segmentation \cite{Berhane2020FullyLearning}. Thus, we believe ML can provide segmentation with ‘real world’ accuracy. Furthermore, there are significant advantages of ML over manual segmentation including very fast segmentation without user interaction and perfect reproducibility, due to its deterministic nature. This makes ML particularly useful for removing clinical bottlenecks and accelerating population-based research.

\subsection{Relationship Between CFD and Segmentation Errors}\label{discussion-cfd}

We demonstrated reasonable agreement in velocity and pressure fields calculated from ML and manual segmentations. Importantly, the differences in CFD metrics using ML vs manual segmentations were of a similar magnitude to those between two independent manual segmentations. This suggests that ML can be successfully used to provide a starting point for CFD simulations, with accuracy similar to inter-observer variability. However, there were some differences in CFD metrics between ML vs GT segmentations, particularly for pressure calculations. We think pressure errors are higher because local deviations in surface geometry tend to cause only local velocity field derangement, but have a global effect on upstream pressures. This can be seen in the worst-case aorta, where a kink in the ground truth descending aorta results in localized flow acceleration, and significantly altered upstream pressures. 

Interestingly, we found no significant correlations between image-based segmentation metrics and errors in the pressure and velocity fields. This suggests that neither overlap-based (Dice, IoU) nor boundary distance-based (HD, ASD) metrics can accurately capture the features that ensure CFD accuracy. This may be because CFD models are highly sensitive to local geometric errors, while segmentation metrics are global and therefore may not fully capture these localized deviations. Another reason may be that differences in clipping (which were not accounted for by segmentation metrics) are responsible for some of the CFD errors, as shown by our analysis of equally clipped data. However, significant CFD errors remained after removing this confounding factor, and these errors were still not correlated with image-based segmentation metrics. Irrespective of the cause, the poor correlation between segmentation and CFD errors has some important implications. Specifically, in our application it might be better to combine conventional global image-based losses with more CFD specific objective measures during training.

Computational fluid dynamics can benefit in several ways from machine learning. Firstly, ML segmentation is completely automated and very fast, enabling significant reduction in pre-processing time, one of the major impediments to clinical uptake. Secondly, ML segmentations are completely reproducible, and this is important as we have shown significant human inter-observer variability. Finally, there has been recent work demonstrating the use of ML to accelerate the CFD simulations. Combined with ML segmentation this would substantially reduce the time taken to perform CFD and make CFD much more attractive for routine clinical use.

\subsection{Limitations}\label{discussion-limitations}

One of the main limitations of this study was that a simplified CFD model was applied across all subjects (laminar, steady state with no patient-specific parameters). This was done to better isolate the effect of segmentation differences on the resulting CFD model. However, it does limit the patient specific aspect of these comparisons and in the future, boundary conditions for each subject (such as velocity profiles taken from phase contrast MRI) could be incorporated into the model. Furthermore, now that we have demonstrated good agreement using simple CFD models, the utility of ML segmentation for more complex CFD models should be investigated.

Another limitation is that the methods used for comparison don’t necessary account for the full flow field. We used plane-averaged pressures and velocities along the length of the centreline to quantitatively compare different CFD models. However, this averaging does lead to a loss of localized details in the flow fields. Additionally, the slice locations were determined independently for ML, GT and SO models, so there may not be an exact one-to-one correspondence. In future studies, particularly if using more complex CFD models, new metrics of CFD errors that capture subtle deviations will need to be developed.

\section{Conclusions}\label{conclusions}

A convolutional neural network was developed, optimized and trained for segmentation of the aorta and the pulmonary arteries in 3D cardiovascular MRI. The segmentation network was validated for its primary purpose: the creation of CFD models and calculation of flow fields. Segmentation errors in terms of Dice, IoU, HD and ASD as well as derived pressure and velocity field errors were in the range of human inter-observer variability. The proposed method could help to automate clinical hemodynamic assessment workflows and improve their robustness.

\newpage
\section{Declarations}\label{declarations}

\subsection{Ethics approval and consent to participate}\label{ethics}

The use of retrospectively collected training and test data was approved by the local research ethics committee, and written consent was obtained from all subjects/guardians (Ref: 06/Q0508/124).

\subsection{Consent for publication}\label{consent}

Written consent to use the data for publication was obtained from all subjects/guardians (Ref: 06/Q0508/124).

\subsection{Availability of data and materials}\label{availability}

The datasets used and/or analysed during the current study are available from the corresponding author on reasonable request.

\subsection{Competing interests}\label{interests}

The authors declare that they have no competing interests.

\subsection{Funding}\label{funding}

This work was supported by UK Research and Innovation (MR/S032290/1), Heart Research UK (RG2661/17/20), the British Heart Foundation (NH/18/1/33511, PG/17/6/32797), the Engineering and Physical Sciences Research Council (EP/N02124X/1) and the European Research Council (ERC-2017-StG-757923).

This report incorporates independent research from the National Institute for Health Research Biomedical Research Centre Funding Scheme. The views expressed in this publication are those of the authors and not necessarily those of the NHS, the National Institute for Health Research or the Department of Health. The funders had no role in study design, collection, analysis and interpretation of the data, decision to publish, or preparation of the manuscript.

\subsection{Authors’ contributions}\label{contributions}

\begin{itemize}
\item JMT contributed to the conception and design of this work, pre-processed the images and labels, designed, implemented, trained and optimized the neural network and related functionality, extracted the reported metrics, interpreted the results, performed the statistical analysis and was a major contributor in drafting the manuscript.
\item EP contributed to the conception and design of this work, designed and implemented the semi-automatic mesh processing pipeline, performed the computational fluid dynamics simulations, extracted the reported pressure and velocity field data, interpreted the results and was a major contributor in drafting the manuscript.
\item RJ performed the reference ground truth segmentation.
\item ES contributed to the conception and design of this work.
\item RP contributed to the preparation of the external test set.
\item AS contributed to the preparation of the external test set.
\item CC contributed to the conception and design of this work.
\item JS contributed to the conception and design of this work and substantially revised the manuscript.
\item SS contributed to the conception and design of this work and substantially revised the manuscript.
\item VM contributed to the conception and design of this work, pre-processed the images and labels, performed the secondary observer segmentation and contributed to draft and substantially revise the manuscript.
\end{itemize}

All authors read and approved the final manuscript.

\subsection{Acknowledgements}\label{acknowledgements}

Not applicable.

\backmatter

\begin{appendices}

\newpage
\section{Sensitivity Analysis: Laminar vs Turbulent}\label{appA}

It was found that an element count of ~300,000 was well-suited to capture the flow details in both the aorta and the PA, and for yielding a stable solution. Iterations of 300 and 750 were sufficient for the aorta and PA respectively, in order to reach convergence and accurate results. The error (MAPE) for the aorta with 315,766 elements vs the highest resolution aorta with 1,102,454 elements was 2.02\% in pressure, and 3.4\% in velocity. The error (MAPE) for the PA with 398,947 elements vs the highest resolution PA with 1,116,831 elements was 3.23\% in pressure, and 2.15\% in velocity. Lastly, the Reynold’s number was computed on both test cases, and resulted in 3221.8 and 4250.5 for the aorta and pulmonary artery respectively. The same simulations were carried out with a k-omega turbulence model (with default parameters), which showed no difference when compared to the laminar model.

\begin{figure}[htbp]
\centering
\includegraphics[width=0.9\textwidth]{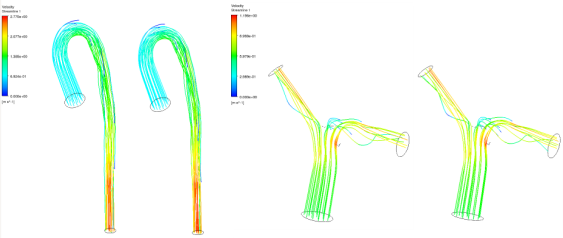}
\caption{Difference velocity fields in Ao and PA when applying a k-omega turbulence model(left = laminar, right = turbulent).}\label{figA1}
\end{figure}

\newpage
\section{Hyperparameter Optimization}\label{appB}

\begin{figure}[htbp]
\centering
\includegraphics[width=0.9\textwidth]{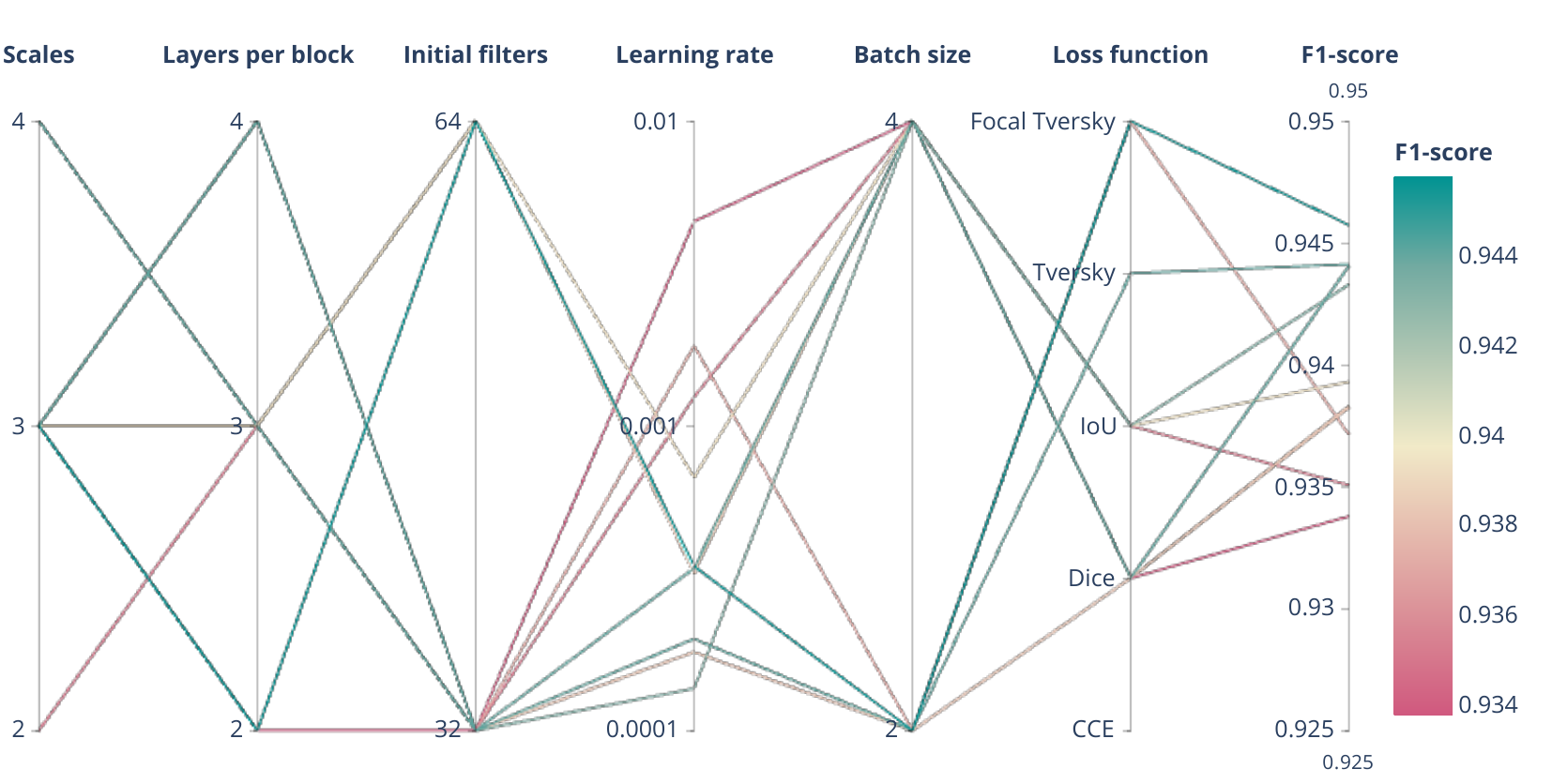}
\caption{Parallel coordinates view of the top 10 hyperparameter configurations. Each hyperparameter as well as the mean validation Dice score is shown on its own axis. Each coloured line represents a hyperparameter combination, with vertices at the corresponding values on the parallel axes.}\label{figB2}
\end{figure}

Training, including hyperparameter optimization, took ~24 h, during which a total of 124 hyperparameter configurations were sampled by the Hyperband iterations. The top 10 performing configurations in terms of mean validation Dice score are reported in Table \ref{tabB1} and Fig. \ref{figB2}.

\begin{table}[htbp]
\centering
\caption{Top 10 hyperparameter configurations.}\label{tabB1}
\begin{tabular}{@{}llp{1.2cm}p{1cm}lp{1cm}ll@{}}
\toprule
\textbf{\#} & \textbf{Scales} & \textbf{Layers per block} & \textbf{Initial filters} & \textbf{Learning rate} & \textbf{Batch size} & \textbf{Loss function} & \textbf{Dice}  \\
\midrule
1  & 3 & 2 & 64 & $3.46 \cdot 10^{-4}$ & 2 & Focal Tversky & 0.946 \\
2  & 3 & 4 & 32 & $1.99 \cdot 10^{-4}$ & 2 & Tversky       & 0.944 \\
3  & 4 & 3 & 32 & $3.40 \cdot 10^{-4}$ & 4 & Dice          & 0.944 \\
4  & 4 & 3 & 32 & $1.37 \cdot 10^{-4}$ & 4 & IoU           & 0.943 \\
5  & 3 & 3 & 64 & $6.77 \cdot 10^{-4}$ & 4 & IoU           & 0.939 \\
6  & 3 & 3 & 64 & $3.27 \cdot 10^{-4}$ & 4 & Dice          & 0.938 \\
7  & 4 & 3 & 32 & $1.81 \cdot 10^{-4}$ & 2 & Dice          & 0.938 \\
8  & 3 & 4 & 32 & $1.83 \cdot 10^{-3}$ & 2 & Focal Tversky & 0.937 \\
9  & 2 & 3 & 32 & $1.24 \cdot 10^{-3}$ & 4 & Jaccard       & 0.935 \\
10 & 3 & 2 & 32 & $4.68 \cdot 10^{-3}$ & 4 & Dice          & 0.934 \\
\botrule
\end{tabular}
\end{table}

The best performing configuration was as follows: scales = 3, layers per block = 2, initial filters = 64, learning rate = $3.46 \cdot 10^{-4}$, batch size = 2, and loss function = focal Tversky. This model was selected and used in all further experiments.

The top 8 performing configurations have 3 or 4 scales, which suggests that models with 2 scales may have too limited representational power or deep layer receptive fields. Models with at least 3 layers per block also tended to perform better, with the notable exception of the best performing model, which had 2. A relatively small learning rate also seems to be advantageous, with 6 of the top 7 configurations below 0.0005. Confusion-based losses seem to outperform cross-entropy, which does not appear on any of the top 10 configurations, but none of those appears to consistently outperform the rest. Finally, although our best model had 64 initial filters, the abundance of top performing models with 32 filters suggests it might be possible to find a 32-filter configuration with little or no performance penalty, which would result in a 2x reduction in the number of parameters and computational cost.

\newpage
\section{Manual and Equal Clipping}\label{appC}

\begin{figure}[htbp]
\centering
\includegraphics[width=0.9\textwidth]{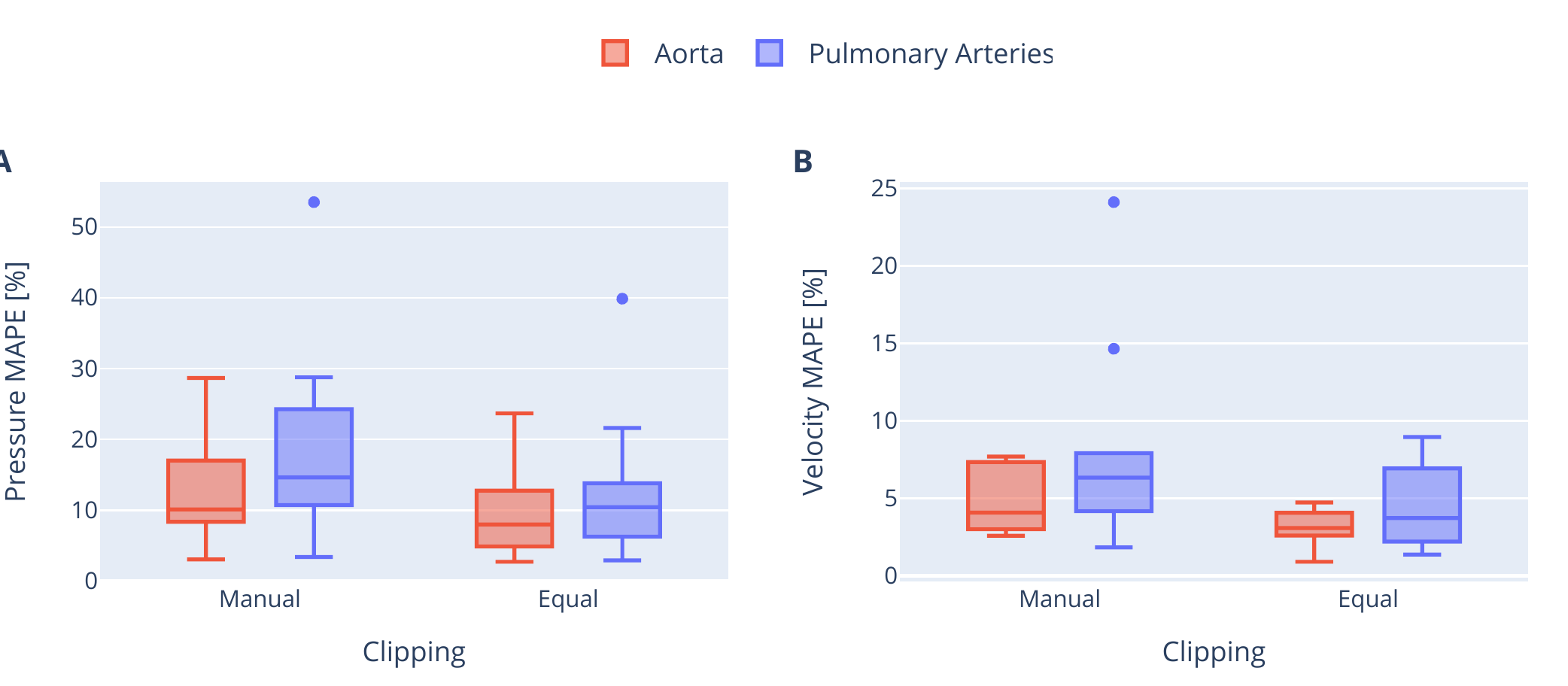}
\caption{Pressure and velocity mean average percentage errors (MAPE) for manual and equally clipped aortas and pulmonary arteries.}\label{figC3}
\end{figure}

In all cases, the equally clipped data produces better agreement between the ML and GT CFD simulations (Fig. \ref{figC3}). The median pressure and velocity error in aortas is reduced through equal clipping by 2.1 and 1.0 percentage points, respectively. The median pressure and velocity error in PAs is reduced through equal clipping by 4.2 and 2.7 percentage points, respectively. The magnitude of outliers is also reduced through equal clipping.

\end{appendices}

\newpage
\bibliography{article}

\end{document}